\documentclass[aps]{revtex4}          
\usepackage{graphicx}
\begin{document}

\title{Quantum Gravity and the Correspondence Principle
}


\author{Ka\'{c}a Bradonji\'{c}}
\affiliation{%
Department of Physics, Boston University, Boston, MA 02215 USA
\begin{center}
kacha@physics.bu.edu
\end{center}
Essay written for the Gravity Research Foundation 2011 Awards \\for Essays on Gravitation
}

\date{Submitted on March 31, 2011}
\maketitle
\section*{Summary}
Standard approaches to quantum gravity start with a pre-spacetime structure and attempt, in accordance with Bohr's correspondence principle, to recover the pseudo-Riemannian manifold in the low energy limit. These approaches assume there is a smooth transition from quantum gravity to general relativity common to successful quantum theories. However, as gravitational field, and hence spacetime, cannot be considered in isolation from physical fields, discontinuities in properties of physical fields, such as loss of mass at the electroweak symmetry breaking scale, may result in change of spacetime structure somewhere between the Planck scale and the scale where general relativity holds true. As a result, the correct theory of quantum gravity may not have general relativity as its low energy limit. 
\newpage

Standard approaches to the problem of quantum gravity start with a pre-spacetime structure, assumed to exist at the Planck scale, and attempt to recover the full pseudo-Riemannian manifold in the low energy limit \cite{Thiemann,Rovelli,Fotini,Sorkin,Loll}.  This limit is expected in accordance with Bohr's correspondence principle which states that the quantum theory must asymptotically approach its classical counterpart in the limit of large quantum number \cite{Messiah}. Successfully applied in the case of quantum mechanics and quantum electrodynamics, correspondence principle is often used as a guide and a test for a potentially successful quantum theory of gravity. All these approaches assume there is a smooth transition from the Planck scale quantum gravity to the general theory of relativity (GR). However, gravitational field is distinct from the other physical fields in that it, and hence spacetime geometry, cannot be considered in isolation from all the other physical fields.  According to Einstein, the special and general theory of relativity, and consequently the attribution of pseudo-Riemannian geometry to spacetime, rest on physically meaningful notions of length, as that which is measured by a rigid rod, and time, as that which which is measured by a physical clock. Einstein goes as far as to say that the entire GR framework rests on the assumption that two line segments defined on rigid bodies equal at some time and place, are equal always and everywhere \cite{Einstein1961}. Assuming that pseudo-Riemannian (or any other) geometry is independent of the physical fields present is not substantiated. As we consider smaller length scales, we may find that spactime structure changes well before we need a quantum description of gravity. In such case, the limiting case of a correct quantum theory may not be GR as we know it.

The first change in the nature of space and time may have been expected to appear at the scales at which notion of a rigid rod breaks down. Einstein himself questioned the applicability of GR to sub-molecular scales and conceded that ``physical interpretation of geometry breaks down when applied immediately to spaces of sub-molecular order of magnitude" \cite{Einstein1922}. At an energy regime where the notion of ``rigid body" becomes meaningless, any concept derived from the concept of a ``rigid body" must be subject to scrutiny. Today we know that relativity is applicable to sub-molecular scales. Although initially defined as measured by a rigid rod and a physical clock, the notions of length and time can be retained at the particle level if they are redefined and inferred from paths of massless and massive particles.  Ehlers, Pirani, and Schild showed that a full pseudo-Riemannian spacetime geometry can be constructed from paths of massless and massive particles by imposing two physically motivated compatibility conditions between the two sets of paths \cite{Ehlers1972,Pirani1973}.

A second natural scale at which a discontinuity in the nature of spacetime may appear is the electroweak symmetry breaking scale (EWSB) \cite{Bradonjic}. According to the Standard Model (SM), prior to EWSB, the Higgs field had a vanishing vacuum expectation value and all the particles were massless \cite{Glashow, Weinberg,Salam}. One may object that particles are never really massless due to radiative corrections which induce thermal mass terms and that, due to this induced mass, particles propagate along non-null geodesics in thermal plasma.  However, the framework of thermal field theory assumes the existence of the full pseudo-Riemannian structure of spacetime. It also requires that one is working at finite temperature, but accessing a short distance scale, in a high energy collision for example, does not necessarily imply that one is working in a thermal background corresponding to that scale. In addition, one can always consider what happens at length scales which are below the mean free path of the collisions with the thermal background. Recognizing the importance of physical justification for mathematical structures used to describe spacetime, we must admit that the use of full pseudo-Riemannian geometry as the accurate description of spacetime geometry when no massive particles are present is unwarranted unless some physical justification is provided. 

A hint to a possible description of spacetime at such a scale may be found in the previously mentioned work by Ehlers, Pirani, and Schild who have shown that a full pseudo-Riemannian geometry of spacetime can be constructed from structures defined by paths of massless and massive particles and by imposing two compatibility conditions between them \cite{Ehlers1972,Pirani1973}. The construction assumes a differentiable manifold and employs light rays (or free massless particles) and free (not under influence of anything but gravitational effects) massive particles as test bodies.  All the considerations of their formalism are local, assume that the manifold and the curves in question are differentiable, and treat particles as classical objects. In broad brushstrokes, the key steps of this axiomatic construction are:

\begin{itemize}
\item {The propagation of light determines at each point of spactime an inÞnitesimal null cone and hence deÞnes a 
\emph{conformal structure} $\mathcal{C}$ on $M$. Light rays are represented by null geodesics which are null curves contained in null hypersurfaces.}
\item{The motions of freely falling massive particles determine a family of preferred unparametrized time-like curves at each point, and such a family at each point of $M$ defines a \emph{projective structure} $\mathcal{P}$ on $M$. World lines of freely falling particles are said to be $\mathcal{C}$-time-like geodesics of $\mathcal{P}$.}
\item{Requiring two compatibility conditions: 1) that the null geodesics are also geodesics of $\mathcal{P}$, and 2) that the ticking rate of a clock is independent of its history, leads to the full pseudo-Riemannian geometry.}
\end{itemize}

While such extrapolation may be possible, one should keep in mind Einstein's warning that ``even when it is a question of describing the electrical elementary particles constituting matter, the attempt may still be made to ascribe physical importance to those concepts of fields that have been physically defined for the purpose of describing the geometrical behavior of bodies that are large as compared with the molecule. Only the outcome can decide the justification of such an attempt, which postulates physical reality for the fundamental principles of Reimann's geometry outside of the domain of their physical definitions" \cite{Einstein1922}. 

If we restrict ourselves to an energy regime where there are no massive particles, the axiomatic construction of the pseudo-Riemannian geometry of Ehlers, Pirani and Schild is not possible because there are no massive particles for the construction of the projective structure. But in Einstein's own words,``[a]ccording to the general theory of relativity, the geometrical properties of space are not independent, but they are determined by matter" \cite{Einstein1961}. Since the projective structure cannot manifest itself in such a ``massless" regime, maintaining that $\mathcal{P}$, and hence the pseudo-Riemannian geometry, remains a property of spacetime is unjustified. We then have to concede that there may be another description of spactime valid between the Planck scale and the scale where GR holds true. If that is the case, a correct quantum theory of gravity, which may be an accurate description at the Planck scale, would not have GR as its classical limit. Instead, its limit would be the theory which accurately describes spacetime, and hence gravitation, right above the EWBS.

The electroweak symmetry breaking scale is interesting, but not the only scale at which a discontinuous shift in spacetime structure may occur. As higher energies are tested experimentally and our existing models are modified to accommodate the empirical observations, we may find that a change in spactime geometry may occur, if at all, at higher energies and due to different physical reasons. Regardless of how and at what scales such change may happen, a possibility that GR may not be the limiting form of a corresponding quantum theory should be taken in consideration in our discussions of what constitutes a good candidate for a correct quantum theory of gravity. 

\begin{acknowledgements}
I would like to thank John Stachel for introducing me to the work on conformal and projective structures and commenting on an earlier draft of this essay.
\end{acknowledgements}



\end{document}